\documentclass[twocolumn]{aa}
\usepackage{graphicx}
\usepackage{txfonts}
\def\kms{\relax \ifmmode {\,\rm km\,s}^{-1}\else \,km\,s$^{-1}$\fi}
\def\ks{\relax \ifmmode  K_{\rm s}\else $K_{\rm s}$\fi}
\def\ha{\relax \ifmmode {\rm H}\alpha\else H$\alpha$\fi}
\def\hb{\relax \ifmmode {\rm H}\beta\else H$\beta$\fi}
\def\hi{\relax \ifmmode {\rm H\,{\sc i}}\else H\,{\sc i}\fi}
\def\hii{\relax \ifmmode {\rm H\,{\sc ii}}\else H\,{\sc ii}\fi}
\def\h2{\relax \ifmmode {\rm H}_2\else H$_2$\fi}
\def\lha{\relax \ifmmode L_{{\rm H}\alpha}\else $L_{{\rm H}\alpha}$\fi}
\def\shi{\relax \ifmmode \sigma_{{\rm HI}}\else $\sigma_{\rm HI}$\fi}
\def\sh2{\relax \ifmmode \sigma_{{\rm H}_2}\else $\sigma_{{\rm H}_2}$\fi}
\def\degr{\hbox{$^\circ$}}
\def\arcmin{\hbox{$^\prime$}}
\def\arcsec{\hbox{$^{\prime\prime}$}}
\def\deg{\hbox{$^\circ$}}

\def\sec{\hbox{$^{\prime\prime}$}}
\def\fdg{\hbox{$.\!\!^\circ$}}
\def\fs{\hbox{$.\!\!^{\rm s}$}}
\def\farcm{\hbox{$.\mkern-4mu^\prime$}}
\def\farcs{\hbox{$.\!\!^{\prime\prime}$}}
\def\degd#1.#2{ #1\fdg#2 }                 
\def\mind#1.#2{ #1\farcm#2 }               
\def\secd#1.#2{ #1\farcs#2 }               
\def\hhh{\ifmmode {\rm ^h}              
         \else {${\rm ^h}$}
         \fi}
\def\sss{\ifmmode {\rm ^s}              
         \else {${\rm ^s}$}
         \fi}
\def\hms#1h#2m#3s{                      
                  \relax
                  \ifmmode #1^{\rm h}\,#2^{\rm m}\,#3^{\rm s}
                  \else \hbox{$#1^{\rm h}\,#2^{\rm m}\,#3^{\rm s}$}
                  \fi
                 }
\def\dms#1d#2m#3s{                      
                  \relax
                  #1\degr\,#2\arcmin\,#3\arcsec 
                 }
\def\hmsd#1h#2m#3.#4s{                  
                      \relax
                      \ifmmode #1^{\rm h}\,#2^{\rm m}\,#3\fs#4
                      \else \hbox{$#1^{\rm h}\,#2^{\rm m}\,#3\fs#4$}
                      \fi
                     }
\def\dmsd#1d#2m#3.#4s{                  
                      \relax
                      #1\degr\,#2\arcmin\,#3\farcs#4
                     }
\def\mag{\relax                          
        \ifmmode ^{\rm m}
        \else $^{\rm m}$
        \fi
       }
\def\magd#1.#2{                          
              \relax
              \ifmmode #1^{\rm m}
                       \hskip-0.55em.\hskip0.22em#2
              \else \hbox{#1$^{\rm m}
                    \hskip-0.55em.\hskip0.22em$#2}
              \fi
             }
\input psfig.sty
\begin{document}
\title{Structure and star formation in disk galaxies III.}
\subtitle{Nuclear and circumnuclear \ha\ emission}
\author{J.~H.~Knapen
}

\offprints{J. H. Knapen}  

\institute{Centre for Astrophysics Research,
University of Hertfordshire, Hatfield, Herts AL10 9AB, U.K.\\
\email{j.knapen@star.herts.ac.uk} }

\date{Received ; accepted 1 Sept. 2004}

\abstract{

From \ha\ images of a carefully selected sample of 57 relatively
large, Northern spiral galaxies with low inclination, we study the
distribution of the \ha\ emission in the circumnuclear and nuclear
regions.  At a resolution of around 100 parsec, we find that the
nuclear \ha\ emission in the sample galaxies is often peaked, and
significantly more often so among AGN host galaxies.  The
circumnuclear \ha\ emission, within a radius of two kpc, is often
patchy in late-type, and absent or in the form of a nuclear ring in
early-type galaxies.  There is no clear correlation of nuclear or
circumnuclear \ha\ morphology with the presence or absence of a bar in
the host galaxy, except for the nuclear rings which occur in barred
hosts. The presence or absence of close bright companion galaxies does
not affect the circumnuclear \ha\ morphology, but their presence does
correlate with a higher fraction of nuclear \ha\ peaks. Nuclear rings
occur in at least 21\% ($\pm$5\%) of spiral galaxies, and occur
predominantly in galaxies also hosting an AGN.  Only two of our 12
nuclear rings occur in a galaxy which is neither an AGN nor a
starburst host.  We confirm that weaker bars host larger nuclear
rings.  The implications of these results on our understanding of the
occurrence and morphology of massive star formation, as well as
non-stellar activity, in the central regions of galaxies are
discussed.

}

\maketitle

%

\section{Introduction}

Observations of the central regions of spiral galaxies reveal a number
of features which are believed to be directly related to the dynamics
of the host galaxy, such as bars, rings, and enhanced star formation
(SF), possibly in the form of a starburst.  Over the past decades, and
more rapidly so in recent years thanks to a deluge of new high-quality
observations and modelling, a general picture has emerged (see, e.g.,
review by Shlosman 1999) where inflowing gas can lead to SF in the
nuclear or circumnuclear regions. Theoretically at least, a connection
between the inflowing gas and non-stellar nuclear activity seems
plausible (e.g., Shlosman et al. 1989).  Observationally, however, it
is proving rather difficult to provide evidence for direct links
between activity and bars or interactions, the main large-scale
mechanisms to cause the loss of angular momentum, and thus to enable
gaseous inflow.  For instance, the bar fraction in Seyfert host
galaxies is higher than that in non-active galaxies, as proven by
several recent studies, but at a formal significance level of only
some 2.5$\sigma$ (Knapen et al.  2000; Laine et al. 2002; Laurikainen
et al. 2004).

Detailed studies of the distribution of the SF in the nuclear and
circumnuclear regions have usually been based on the study of small
numbers of objects, and where larger samples have been used, these
have mostly not been selected without different biasing factors in the
sample selection (e.g., the comprehensive study by Buta \& Crocker
1993 included nuclear rings found by targeting galaxies which were
known to host inner and/or outer rings).  In particular, no unbiased
statistical studies of star-forming nuclear rings have been published.
This is the main justification for the use, in the present paper, of
our data set, which lends itself to a statistical study of the nuclear
and circumnuclear SF even though in itself the images are not nearly
as spectacular as, say, {\it Hubble Space Telescope (HST)} images. Our
data set consists of a total of 57 continuum-subtracted images of
spiral galaxies in the \ha\ line. \ha\ directly traces massive SF,
with a possible contribution from AGN emission in the nuclear regions
of active galaxies.

As noted by Buta \& Combes (1996) in their review, it is relatively
difficult to identify and study nuclear rings, given the large amount
of background light from the bulge or inner regions of the disk. This
means that only prominent nuclear rings, which can be considered
rather extreme in size and/or relative brightness, can be found from
the plates on which major galaxy catalogues and classifications are
based.  For a more complete statistical study, surveys are needed in
wavebands which specifically trace nuclear rings. Capitalising on the
enhanced massive SF in star-forming nuclear rings, the \ha\ line, as
used in the current data set, is one of the best tracers available.

From our set of \ha\ images, at a typical spatial resolution of around
a hundred parsec, we classify the nuclear and circumnuclear \ha\
morphology into a small number of distinct categories.  This is very
similar in principle to work by Pogge (1989a,b).  The current work is
more than a consistency check, though, and adds further value by the
use of data of slightly higher angular resolution, as well as a more
thoroughly applied set of criteria, for instance on the spatial extent
of the nuclear and circumnuclear regions.

After a brief summary of the properties of the data set (Sect.~2) and
an overview of the procedures and terminology used (Sect.~3), we
discuss the results in terms of the properties of the host galaxies in
Sect.~4. We pay attention to the nuclear rings and their hosts in
Sect.~5, discuss possible relations to nuclear activity in Sect.~6,
and summarise our findings and conclusions in Sect.~7.

\section{Sample selection and observations}

The selection criteria for the sample of 57 spiral galaxies, as well
as some of its statistical properties, have been described by Knapen
et al. (2003, hereafter Paper~I).  Basically, we selected all galaxies
which are larger than 4.2~arcmin in diameter, of spiral type, inclined
less than 50\deg, and visible from the Northern hemisphere
($\delta>-20$\deg).  The resulting sample contains spiral galaxies of
all morphological types, with and without bars, with and without
nuclear activity, and of all spiral arm classes (cf.  Elmegreen \&
Elmegreen 1987).  All sample galaxies are relatively nearby, with an
angular scale of 82 parsec/arcsec in the median, and of at most 183
parsec/arcsec (in the case of NGC~5371). For the median value, a
two~kpc radius is equivalent to just over 24 arcsec, or 100 pixels.

In this paper, we use a set of continuum-subtracted \ha\ images of our
sample of 57 galaxies, as described by Knapen et al. (2004a, hereafter
Paper~II). The \ha\ line observations were obtained mostly with the
1~m Jacobus Kapteyn Telescope (JKT) on La Palma.  Images of several,
mostly larger, galaxies were obtained with the 2.5~m Isaac Newton
Telescope or the 4.2~m William Herschel Telescope (WHT), as detailed
in Paper~II. Continuum emission was subtracted using images in the $R$
or $I$ broad bands or as obtained though a medium-width continuum
filter (see Paper~II).  The spatial resolution of the \ha\ images is
reasonable with a median value of 1.54~arcsec, well sampled with the
original pixel size of 0.331~arcsec (smaller for those images obtained
with the WHT), re-sampled in the data reduction process to
0.241~arcsec (see Paper~II for details).  Whereas the resolution is
not ideal for studying circumnuclear emission in detail, this data set
does allow us to compare the circumnuclear and nuclear \ha\ morphology
across a wide range of spiral types.

In Paper~II, we discuss the uncertainties resulting from the use of
different filters in the continuum subtraction.  We show there that
these uncertainties are generally small, except possibly in the
nuclear regions, and especially in the few cases where the $I$-band
filter has been used for the determination of the continuum.  This is
mostly because the $I$ filter transmits at longer wavelengths than the
\ha\ line, and hence dust extinction can influence the result. We can
directly check the impact of dust by comparing our results for the
nuclear emission with those obtained by B\"oker et al.  (1999) from
Pa$\alpha$ images, obtained at NIR wavelengths where the effects of
extinction are much reduced.  The overlap between their and our
samples is only five galaxies, but the nuclear classifications for
those agree in all cases: one galaxy (NGC~628) is classified as having
no nuclear emission, four others (NGC 3184, 4395, 5474 and 6946) as
having a nuclear peak.  Of these five, we used the $I$-band filter to
subtract the continuum in the case of NGC~628, and the $R$-band for
the remaining four galaxies.  This test, although unfortunately based
on a handful of galaxies only, does show that the classification of
nuclear emission as peaked is not systematically affected by the use
of broad-band images as continuum.

In a few cases (those of NGC~488, NGC~1169, NGC~3368, NGC~4725 and
NGC~6384) the continuum subtraction is less than perfect, and prevents
any reliable determination of the nuclear \ha\ morphology. In
addition, in the case NGC~7741, it is not possible to pinpoint the
location of the galaxy nucleus (even in the NIR image of Paper~I the
nuclear region has multiple emission peaks). The nuclear \ha\
morphology has been tabulated as ``N/A'' (Table~\ref{resultstab}) in
these six cases, and these six galaxies have not been included in the
statistical analysis, below, of nuclear emission.

\begin{table*}
\centering
\begin{tabular}{lcccccccc}
\hline
NGC & Type & Activity & Nucl. \ha& Circ. Nucl. & Nucl. \ha & Circ. Nucl. & Ring & Ring \\
 & & & structure & structure & structure & structure & diam. & diam. \\
 & (RC3) & (NED) & (this paper) & (this paper) & (Pogge) & (Pogge) & (arcsec) & (kpc) \\
\hline
210 & .SXS3 & &  peak & patchy & stellar & diffuse bar &  &  \\
337A & .SXS8 & & none & patchy & - & - &  &  \\
488 & .SAR3 & & N/A & none & stellar & patchy &  &  \\
628 & .SAS5 & & none & patchy & none & patchy &  &  \\
864 & .SXT5 & & peak & none & stellar & patchy bar &  &  \\
1042 & .SXT6 & & peak & none & stellar & none &  &  \\
1068 & RSAT3 & Sy1 Sy2 & peak+amorph & ring & - & - & 46 & 3.2 \\
1073 & .SBT5 & & amorph & patchy & amorph & patchy bar &  &  \\
1169 & .SXR3 & & N/A & none & - & - &  &  \\
1179 & .SXR6 & & none & patchy & - & - &  &  \\
1300 & .SBT4 & & peak & ring & ring & none & 10 & 0.9 \\
2775 & .SAR2 & & peak & diffuse & none & none & & \\
2805 & .SXT7 & & none & patchy & - & - &  &  \\
2985 & PSAT2 & LINER & peak & patchy & - & - & & \\
3184 & .SXT6 & HII & peak+amorph & f.ring & stellar & v.f.patchy & 82 & 3.4 \\
3227 & .SXS1P & Sy1.5 & peak+amorph & patchy (bar) & - & - &  &  \\
3344 & RSXR4 & & peak & ring & - & - & 67 & 2.0 \\
3351 & .SBR3 & HII Sbrst & f.peak & ring & ring & none & 20 & 0.8 \\
3368 & .SXT2 & Sy LINER & N/A & f.diffuse & stellar & patchy & & \\
3486 & .SXR5 & Sy2 & peak & ring & - & - & 54 & 1.9 \\
3631 & .SAS5 & & none & patchy & - & - &  &  \\
3726 & .SXR5 & & peak+amorph & patchy & stellar & knot 3.4SW &  &  \\
3810 & .SAT5 & & peak & patchy & amorph & patchy &  &  \\
4030 & .SAS4 & & amorph & patchy & - & - &  &  \\
4051 & .SXT4 & Sy1.5 & stellar+amorph & f.patchy & - & - &  &  \\
4123 & .SBR5 & Sbrst HII & peak & patchy & stellar & f.patchy bar &  &  \\
4145 & .SXT7 & HII/LINER & amorph & patchy & none & f.patchy &  &  \\
4151 & PSXT2 & Sy1.5 & peak & patchy & - & - &  &  \\
4242 & .SXS8 & & none & patchy & - & - &  &  \\
4254 & .SAS5 & & none & patchy & ring & patchy & & \\
4303 & .SXT4 & HII Sy2 & peak & ring & ring & none & 8 & 0.6 \\
4314 & .SBT1 & LINER & peak & ring & ring & none & 16 & 0.8 \\
4321 & .SXS4 & LINER HII & peak & ring & ring & patchy arms & 22 & 1.8 \\
4395 & .SAS9 & LINER Sy1.8 & peak & patchy & - & - &  &  \\
4450 & .SAS2 & LINER & peak & patchy & stellar & none &  &  \\
4487 & .SXT6 & & peak+amorph & patchy & - & - &  &  \\
4535 & .SXS5 & & peak & diffuse & stellar & diffuse &  &  \\
4548 & .SBT3 & LINER Sy & peak & none & stellar & v.f.patchy &  &  \\
4579 & .SXT3 & LINER Sy1.9 & peak & patchy & stellar & v.complex &  &  \\
4618 & .SBT9 & HII & peak & patchy & - & - &  &  \\
4689 & .SAT4 & & none & patchy & none & patchy &  &  \\
4725 & .SXR2P & Sy2 & N/A & none & - & - & & \\
4736 & RSAR2 & Sy2 LINER & peak & ring & stellar & diffuse & 102 & 2.1 \\
5247 & .SAS4 & & none & patchy & none & patchy &  &  \\
5248 & .SXT4 & Sy2 HII & amorph & ring & ring & patchy & 17 & 1.9 \\
5334 & .SBT5 & & none & f.patchy & - & - &  &  \\
5371 & .SXT4 & LINER & peak & ring & - & - & 5 & 1.0 \\
5457 & .SXT6 & & peak+amorph & patchy & - & - &  &  \\
5474 & .SAS6P & HII & peak & patchy & - & - &  &  \\
5850 & .SBR3 & & peak+amorph & none & diffuse & none & & \\
5921 & .SBR4 & LINER & peak & none & stellar & none &  &  \\
5964 & .SBT7 & & none & patchy & - & - &  &  \\
6140 & .SBS6P & & amorph & patchy & - & - &  &  \\
6384 & .SXR4 & LINER & N/A & none & stellar & f.patchy bar &  &  \\
6946 & .SXT6 & HII & peak & patchy & - & - &  &  \\
7727 & SXS1P & & peak+amorph & none & - & - &  &  \\
7741 & .SBS6 & & N/A & patchy & none & patchy bar &  &  \\
\hline
\end{tabular}
\caption{Results of the morphological classification of the nuclear
and circumnuclear \ha\ structure of the 57 sample galaxies. NGC number
(column~1); morphological type (from de Vaucouleurs et al.~1991,
hereafter RC3; column~2); nuclear activity (from the NED, column~3);
classification of nuclear (only the nuclear source) and circumnuclear
(2~kpc radius) structure (this paper, columns~4 and 5, respectively);
similar classifications by Pogge (1989b; columns~6 and 7); and
estimated major axis diameter in arcsec and kiloparsec of nuclear
rings (columns~8 and 9).  }
\label{resultstab}
\end{table*}

\section{Procedure, terminology and results}

We used the Tully (1988) distances to the galaxies to calculate which
region in each image corresponds to a radius of 2~kpc, the radius
taken here to delimit the circumnuclear region.  Using standard image
display methods, the nuclear classification was derived from the
nuclear \ha\ emission (if any) on the smallest measurable scale,
corresponding to the central one to a few seeing elements.  As
mentioned in the previous Section, this corresponds to at most a few
hundred parsec.

Before explaining the various categories used in classifying the
central \ha\ morphology, we will briefly mention a few differences
between the terminology used by Pogge (1989b) and the similar one used
here.  First, Pogge used the term ``stellar'' to describe centrally
peaked emission, but we found only one case (namely that of NGC~4051)
where the nuclear emission peak was not significantly more extended
than the point spread function in the image, as measured from stars in
the original image, before continuum subtraction (and listed in
table~1 of Paper~II). Thus, we prefer to use the term ``peak'' in all
remaining cases of centrally peaked \ha\ emission.  Second, we have
tried to be more consistent in the use of the terms ``nuclear'' and
``circumnuclear''.  In this paper, the latter refers to an area of
2~kpc radius around the nucleus, whereas the former refers to the very
innermost region, of at most a few hundred parsec across.  In
contrast, Pogge (1989b) placed some nuclear rings in the ``nuclear''
emission list, whereas all such rings would be listed as
``circumnuclear'' by us.

That said, the overall classification scheme is very similar to the
one used by Pogge (1989b).  For the nuclear emission, apart from the
{\it stellar} and {\it peak} categories described above, we use the
terms {\it none} to describe, obviously, the absence of any \ha\
emission from the nucleus; {\it amorph} which indicates amorphously
distributed \ha\ emission, without clearly identifiable peaks; and
{\it patchy}, to describe individual regions of \ha\ emission, not
centrally peaked, and resembling \hii\ regions as known in the disks
of galaxies.  In some cases, a peaked nuclear \ha\ distribution is
seen surrounded by amorphous emission which clearly extends beyond the
nuclear peak.  These cases have been marked ``peak$+$amorph''.

The circumnuclear \ha\ emission is classified as follows. {\it None}
implies no discernible \ha\ emission at all within the circumnuclear
area (except possibly in the nucleus); {\it diffuse} means that \ha\
emission is present, but it cannot easily be ascribed to individual
and well-defined \hii\ regions; {\it patchy} is used for galaxies
where well-defined patches of \ha\ emission are seen, caused by
individual \hii\ regions, but not organised in a ring-like fashion;
and, finally, {\it ring}, describes a situation where individual \hii\
regions are organised into a well-defined nuclear ring or pseudo-ring.
In a few cases, we use a rather qualitative classifier such as
``f.patchy'' to indicate the particularly faint nature of the
circumnuclear emission.

In Table~\ref{resultstab} we list our classifications of the nuclear
and circumnuclear \ha\ emission, as well as, for comparison, those
arrived at by Pogge (1989b) for the galaxies in common between his and
our samples (those of our galaxies which were not in his sample are
simply listed as ``--'' in the ``Pogge'' columns of our Table).
Table~\ref{resultstab} also lists the morphological type of the sample
galaxies obtained from the RC3, a description of the nuclear activity
as obtained from the NASA/IPAC Extragalactic Database, NED, and the
diameters of the nuclear rings identified in our study, as measured
from the \ha\ images.

Our results will be analysed below with the aid of tables and figures
showing the \ha\ morphology for different host galaxy types and types
of nuclear activity, but here we just mention that in general, and
taking into account the two specific cases of different use of
terminology by Pogge and this study, his and our classifications agree
very well.  Where they do not agree, the differences can usually be
sourced to our strict use of the 2~kpc radius to limit the area of
circumnuclear emission.

\begin{table*}
\centering
\begin{tabular}{lccccccccccccc}
\hline
Emission & $N$ & \multicolumn{9}{c}{Morphological type $T$} & \multicolumn{3}{c}{Bar type}\\
 & & 1 & 2 & 3 & 4 & 5 & 6 & 7 & 8 & 9 &                   SA & SX & SB\\
\hline
\multicolumn{14}{c}{Total sample}\\
 & 57 & 3 & 7 & 8 & 12 & 11 & 9 & 3 & 2 & 2 &              15 & 29 & 13\\
\multicolumn{14}{c}{Nuclear emission}\\
Peak & 26 & 1 & 5 & 4 & 6 & 5 & 3 & 0 & 0 & 2 &            7 & 12 & 7\\
Peak $+$ amorph & 8 & 2 & 0 & 2 & 0 & 1 & 3 & 0 & 0 & 0 &    1 & 6 & 1\\
Total peak$^1$ & 34 & 3 & 5 & 6 & 6 & 6 & 6 & 0 & 0 & 2 &  8 & 18 & 8\\
None & 11 & 0 & 0 & 0 & 2 & 4 & 1 & 2 & 2 & 0 &            5 & 4 & 2\\
Amorph & 5 & 0 & 0 & 0 & 2 & 1 & 1 & 1 & 0 & 0 &           1 & 2 & 2\\
Stellar + amorph & 1 & & & & & & & & & & & & \\
N/A$^2$ & 6 & & & & & & & & & & & & \\
\multicolumn{14}{c}{Circumnuclear emission}\\
Patchy & 32 & 1 & 3 & 2 & 4 & 8 & 7 & 3 & 2 & 2 &          11 & 14 & 7\\
None & 10 & 1 & 1 & 4 & 2 & 1 & 1 & 0 & 0 & 0 &            1 & 6 & 3\\
Ring & 12 & 1 & 1 & 2 & 6 & 1 & 1 & 0 & 0 & 0 &            2$^3$ & 7 & 3\\
Diffuse & 3 & 0 & 2 & 0 & 0 & 1 & 0 & 0 & 0 & 0 &          1 & 2 & 0\\
\hline
\end{tabular}
\caption{Morphological type distribution and bar presence for the
various types of nuclear and circumnuclear \ha\ emission.  For the
total sample and for nuclear and circumnuclear emission separately,
column~2 gives the total number of galaxies in each category (as
identified in column~1), with the number of galaxies of each
morphological type (from the RC3) in columns 3-11, and bar type (from
the RC3) in columns 12-14.  Notes: $^1$ ``Total peak'' category is the
total of the ``Peak'' and ``Peak + amorph'' categories; $^2$ N/A
indicates that due to difficulties in the continuum subtraction or
determination of the location of the nucleus, no nuclear morphology
could be determined; $^3$ these two galaxies, NGC~1068 and 4736, show
evidence for a small bar from NIR images (see Sect.~5.2).}
\label{typetab}
\end{table*}

As summarised in Table~\ref{typetab}, of the 51 galaxies for which the
nuclear morphology could be classified, 34 (67\%$\pm$7\%, where the
uncertainties quoted are Poisson errors, cf. Laine et al. 2002) show a
peak (of which 8, or 16\%$\pm$5\% of the total, also show amorphous
emission around their peak), 11 (22\%$\pm$6\%) show no emission at
all, five (10\%$\pm$6\%) are classed ``amorph'', and the category
``stellar+amorph'' is used for one galaxy (NGC~4051; 2\%$\pm$3\%).
The circumnuclear \ha\ emission is patchy in a total of 32
(56\%$\pm$7\%) of the 57 sample galaxies, diffuse in three
(5\%$\pm$3\%), in the form of a nuclear ring in 12 (21\%$\pm$5\%), and
absent in 10 (18\%$\pm$5\%) galaxies.

\section{Host galaxy properties}

\subsection{Absolute magnitude, diameter, and distance}

\begin{table}
\centering
\begin{tabular}{lcccccc}
\hline
Emission & $N$ & $D$ & $d$ & $M_B$ & \multicolumn{2}{c}{Companions}\\
& & (kpc) & (Mpc) & (mag) & Yes (\%) & No (\%)\\
\hline
\multicolumn{7}{c}{Nuclear emission}\\
Peak & 26 & 23.4 & 16.8 & $-$20.0 & 64 & 36\\
Peak$+$am. & 8 & 24.0 & 17.0 & $-$20.3 & 18 & 14\\
Total peak & 34 & 23.7 & 16.8 & $-$20.1 & 82 & 50\\
None & 11 & 26.6 & 21.2 & $-$20.0 & 7 & 41\\
Amorph & 5 & 23.0 & 20.7 & $-$19.7 & 11 & 9\\
\multicolumn{7}{c}{Circumnuclear emission}\\
Patchy & 32 & 23.2 & 17.0 & $-$20.0 & 57 & 56\\
None & 10 & 35.6 & 24.3 & $-$20.4 & 13 & 22\\
Ring & 12 & 24.9 & 12.1 & $-$19.8 & 23 & 19\\
Diffuse & 3 & 23.2 & 16.8 & $-$20.1 & 7 & 4\\
\hline
\end{tabular}
\caption{Median values of the host galaxies' diameter (in kpc, from
the RC3; column~3), distance (in Mpc, from Tully 1988; column~4) and
absolute magnitude (calculated using $m_B$ from the RC3 and the
distance; column~5), for the different classes of nuclear and
circumnuclear emission (column~1; the number of galaxies in each
category is shown in column~2). The last two columns show the
fractions of the different nuclear and circumnuclear \ha\ emission
morphologies among those galaxies which have close bright companions
(column~6) and those which do not (column~7). There are 30 and 27
galaxies in these categories, respectively. See Sect.~4.4 for
definitions and further details.}
\label{hosttab}
\end{table}

From the data presented in Table~\ref{resultstab} and the properties
of the sample galaxies (see Paper~I) we have derived information on
how the different classes of nuclear and circumnuclear \ha\ morphology
correlate with the host galaxy parameters diameter, distance, and
absolute magnitude (Table~\ref{hosttab}).

Considering first the nuclear emission, the only trend is that those
galaxies which host a nuclear \ha\ peak are on average closer than
those which don't host a nuclear peak. This is probably simply due to
lower spatial resolution preventing the detection of nuclear
peaks. The fact that the peak host galaxies are not larger or brighter
than the others must be due to the way in which our sample was
constructed, using a lower limit to galaxy diameter.

The distribution of the circumnuclear emission classes shows two clear
effects. First, those galaxies which show no circumnuclear \ha\
emission are in the median further away, brighter, and considerably
larger than other galaxies in the sample. This can be explained as a
selection effect simply because any \ha\ emission is more likely to go
undetected in the more distant galaxies, due to decreasing brightness
and spatial resolution. The most distant galaxies must have the
largest diameters to make it into our sample (we imposed a lower limit
of 4.2 arcmin). Second, the galaxies hosting nuclear rings are
significantly closer (and slightly less bright) in the median than
other galaxies. The most attractive explanation is again that this is
due to a selection effect: namely that nuclear rings can only be
detected in nearby galaxies in a survey such as ours, with limited
spatial resolution. This might imply that there are more nuclear
rings, but unresolved, among the rest of our sample, and hence that
the nuclear ring fraction derived in this paper is really a lower
limit. Only observations of galaxies at higher spatial resolution
could confirm this, but UV imaging with the {\it HST} has failed to
detect a population of such smaller rings (Maoz et al. 1996a, see also
Sect.~5.1).

\subsection{Morphological type of the host galaxy}

\begin{figure}
\psfig{figure=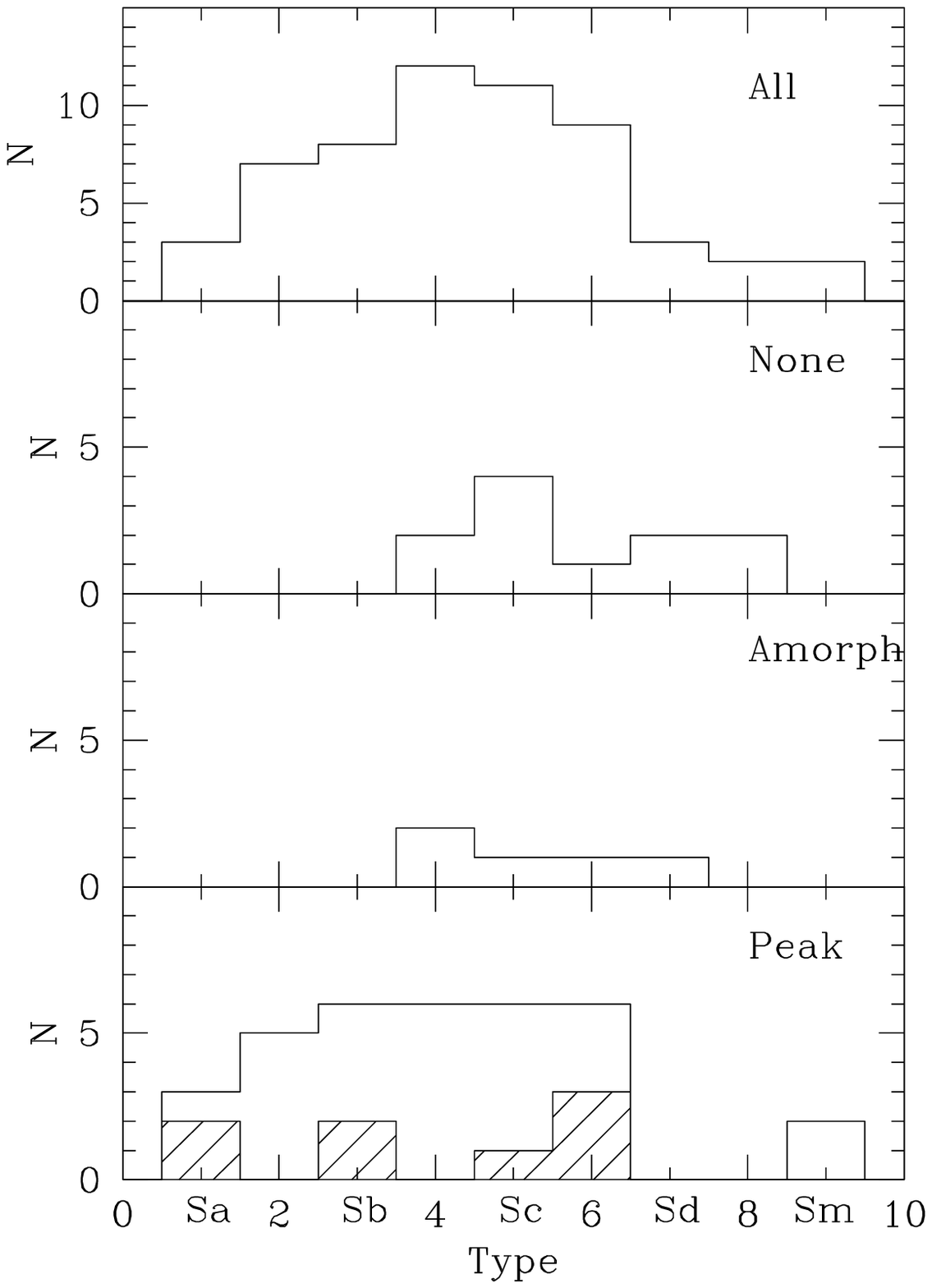,angle=0,width=9cm}
\caption{Set of histograms showing the distribution of nuclear \ha\
emission morphology with host galaxy type, the latter obtained from
the RC3. The different categories of emission are those listed in
Table~\ref{resultstab}.  The ``peak'' histogram refers to the total
numbers of galaxies classified as such, but the subset of
``peak+amorph'' is indicated as the shaded part of the
histogram. Seven galaxies for which either no information can be
obtained, or which are alone in their class, have not been plotted
separately but have been included in the top panel.}
\label{typenucl}
\end{figure}

\begin{figure}
\psfig{figure=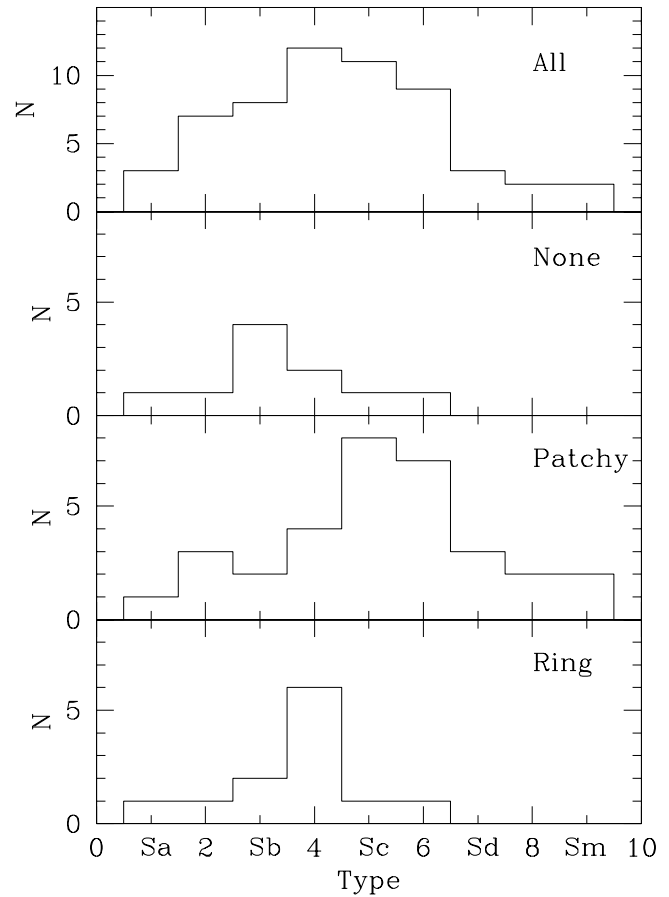,angle=0,width=9cm}
\caption{As   Fig.~\ref{typenucl},  now   for   the  distribution   of
circumnuclear  \ha\ emission  morphology. The  three  ``diffuse'' cases
have not been plotted as a separate histogram.}
\label{typecircumnucl}
\end{figure}

\begin{figure}
\psfig{figure=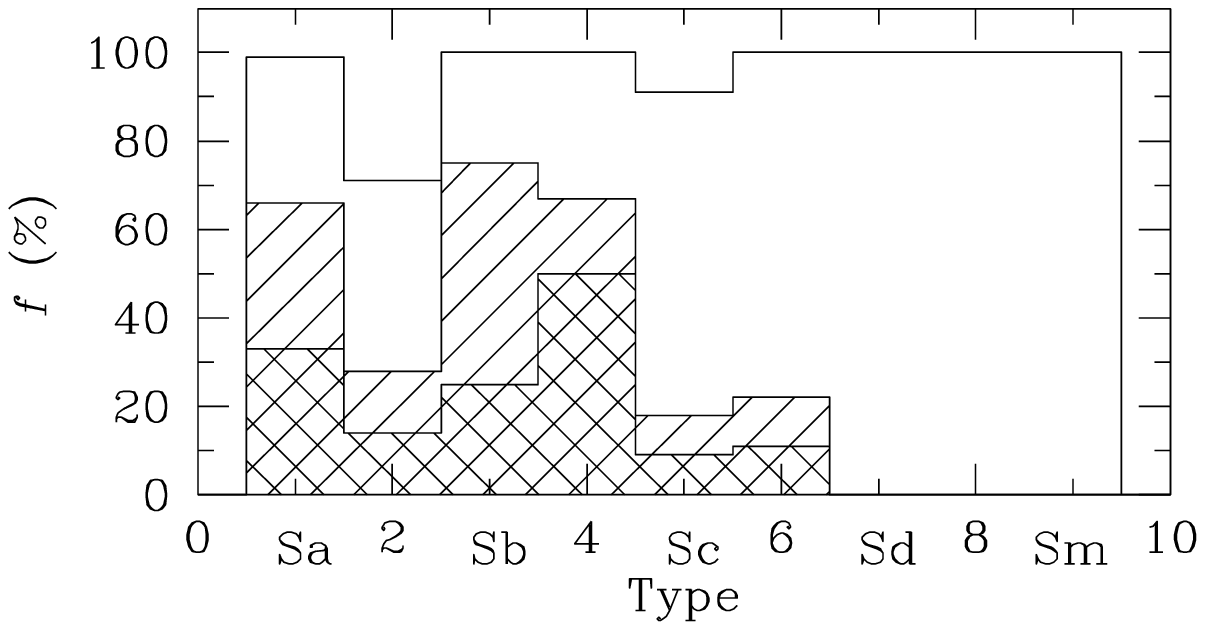,angle=0,width=9cm}
\caption{Histogram showing the cumulative fractions of galaxies of
each morphological type which have ``ring'' (crossed area), ``none''
(hatched), or ``patchy'' (unmarked, top, parts of bars) circumnuclear
morphology.  The fractions do not add up to 100\% for morphological
types 2 and 5 because of the three galaxies with ``diffuse'' \ha\
emission which are not included in this graph.}
\label{typeRatioCN}
\end{figure}

The distribution of nuclear  and circumnuclear \ha\ emission among the
different morphological types of  the sample galaxies is summarised in
Table~\ref{typetab},  and  shown  graphically  in  Fig.~\ref{typenucl}
(nuclear     morphology)     and    Figs.~\ref{typecircumnucl}     and
\ref{typeRatioCN}  (circumnuclear).   As  discussed in  Paper~II,  our
sample spans  the complete range of  spiral types, from Sa  to Sm, but
peaks near types 4 and 5, Sbc and Sc.

As Fig.~\ref{typenucl} shows, a peaked nuclear morphology is the most
common. No trends with morphological type are obvious, and further
analysis is prevented by the small numbers of galaxies in the ``none''
and ``amorph'' classes, or with later morphological types.

The results on circumnuclear \ha\ emission morphology (shown as number
histograms in Fig.~\ref{typecircumnucl}, and as histograms of
fractions of each morphological type which have certain circumnuclear
\ha\ morphology in Fig.~\ref{typeRatioCN}) are more interesting.
First, we find that ``patchy'' emission occurs preferentially in the
later-type galaxies, with practically all galaxies of type Sc and
later in this class (all of those with type Sd or later;
Fig.~\ref{typeRatioCN}).  Second, an absence of circumnuclear emission
(class ``none'') is most often found in the earlier types. Third, the
nuclear ring distribution shows a sharp peak at type 4 (Sbc) and
indicates that nuclear rings are absent from late-type galaxies.  This
will be discussed in more detail below (Sect.~5.1).

The ``patchy'' class traces individual patches of \ha\ emission, as
caused by \hii\ regions, and since these occur more in late- than in
early-type galaxies (e.g., Kennicutt et al.  1989; review by Kennicutt
1998), the result found here is as expected.  Similarly, the ``none''
preference for early-type hosts can be easily understood from the
relative paucity of gas, massive SF, and \hii\ regions in those
galaxies.  But even within this class, some early-type galaxies,
presumably the gas-rich ones, show higher fractions of star-forming
regions (e.g., Pogge \& Eskridge 1987, 1993).  Indeed, we do find that
significant fractions of even the earliest galaxies in our sample show
``patchy'' or ring-shaped \ha\ circumnuclear morphology
(Figs.~\ref{typecircumnucl},~\ref{typeRatioCN}).

\subsection{Bars}

Table~\ref{typetab} shows the bar presence in the sample galaxies, as
derived directly from the RC3 morphological classification. The ``SA''
galaxies are unbarred, and we consider the ``SX'' and ``SB'' galaxies
as barred.  Of the whole sample of 57, 42 galaxies (74\%$\pm$6\%) are
barred, roughly as expected (see, e.g., Mulchaey \& Regan 1997; Knapen
et al.  2000; Eskridge et al.  2000). Apart from the link between bars
and nuclear rings, which will be discussed in more detail below
(Sect.~5), and taking the margins of uncertainty into account,
Table~\ref{typetab} shows no significant effects. Of the 11 galaxies
with no nuclear \ha\ emission, class ``none'', five have no bar (class
SA) which is a slightly higher fraction than expected. It is tempting
to suggest that this is a result of the expectation that bars will
stimulate the central concentration of gas, which can lead to SF and
\ha\ emission, a picture which would qualitatively fit the numbers
observed, but a much enlarged sample with proper NIR imaging to derive
the bar presence is needed before such conclusions can be drawn
reliably.

\subsection{Presence of companion galaxies}

The last two columns of Table~\ref{hosttab} list the results of an
analysis of the presence of close bright companion galaxies, performed
using the HyperLEDA database, hosted by the Observatoire de Lyon
(Paturel et al. 2003). Our criteria for deciding whether a galaxy has
such a companion have been adapted from the works of Schmitt (2001)
and Laine et al. (2002), as follows.  We consider that a sample galaxy
has a close and bright companion if at least one companion galaxy
exists {\it either} within a cylindrical volume of radius 400 kpc and
``depth'' of $\pm500$ km\,s$^{-1}$ in $cz$, centred on the sample
galaxy, and which is not fainter than the sample galaxy by $\Delta
B_T=1.5$ mag, {\it or} within a volume with radius of five times the
optical diameter of the galaxy, $\Delta cz=1000$ km\,s$^{-1}$, and
$\Delta B_T<3$ mag. We find that 30 of our 57 sample galaxies have a
close bright companion, although very few of these are in fact
strongly interacting or in the process of a merger, but all may well
be gravitationally influenced by the companion. The remaining 27 have
no close bright companion, although 21 of these do have a weaker
companion (fainter by more than $\Delta B_T=1.5$ mag) within the
volume of $R=400$~kpc, $\Delta cz=500$ km\,s$^{-1}$.

In Table~\ref{hosttab}, we list the fractions of galaxies which have a
close bright companion following the criteria described above, and of
those which do not.  This analysis shows that, first, galaxies with
close bright companions have more often a nuclear \ha\ peak, and less
often an absence of nuclear \ha\ emission, and, second, that whether a
galaxy has a close bright companion or not does not have a significant
effect on the distribution of the circumnuclear \ha\ emission. The
former result can be interpreted qualitatively in terms of central gas
concentration induced by the interaction between the galaxies (e.g.,
Mihos \& Hernquist 1996, who studied the case of mergers of disk
galaxies of comparable mass), but the absence of any effect on the
circumnuclear star formation is slightly puzzling in this respect.

\section{Nuclear rings}

\subsection{General statistics and host galaxy morphology}

Of our 57 sample galaxies, 12 host a nuclear ring, or
21\%$\pm$5\%. Since we did not make specific assumptions in our sample
selection which might bias the detection of nuclear rings, we can thus
state that about {\it a fifth of disk galaxies host nuclear rings},
which, as far as we are aware, is the first {\it a priori}
determination of the nuclear ring fraction in the general population
of nearby spiral galaxies. Given the data set used in this paper, we
should strictly speak about the fraction of star-forming nuclear
rings, which may be a lower limit to the true nuclear ring fraction
(e.g., Erwin \& Sparke 2001 discuss a category of nuclear rings which
do not currently form massive stars and are identified on the basis of
their broad-band optical colour appearance; these authors report a
total nuclear ring fraction of a third among their sample of
early-type, S0-Sa, galaxies).

Maoz et al. (1996a) observed around 100 nearby galaxies in the UV with
the {\it HST}, and found that only five of these galaxies had a
nuclear ring. As recognised by the authors (Maoz et al. 1996b) this is
mostly due to the small field of view of their images, of only 22
arcsec. For comparison, six or seven of the 12 nuclear rings listed in
Table~\ref{resultstab} are larger than that. What is perhaps more
surprising is that Maoz and collaborators did not find more small
nuclear rings: four of their five were known from ground-based work,
the fifth was a small nuclear ring with a diameter of four arcsec.

\begin{figure}
\psfig{figure=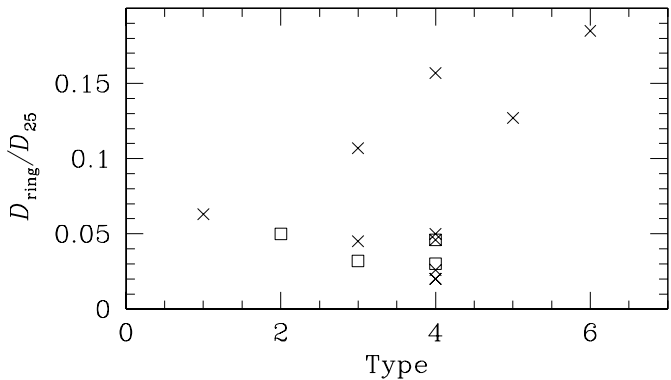,angle=0,width=9cm}
\caption{Relative size,  or ring diameter divided  by galaxy diameter,
of the nuclear rings identified  in our sample (crosses) as a function
of  the morphological  type of  the host  galaxy.  Four  galaxies from
Knapen et  al. (2002) which  are not in  our current sample  have also
been plotted (squares).}
\label{ringtype}
\end{figure}

Nevertheless, our nuclear ring fraction of 21\% is a lower limit due
to the relatively low angular resolution of our \ha\ images. This
limits the radii of the nuclear rings found in our study to a few
hundred parsec, which implies that small nuclear rings such as those
found in NGC~5248 by Laine et al.  (2001) and Maoz et al. (2001), of
around two arcsec (220~pc) in diameter, or in IC~342 (B\"oker et al.
1997), of less than 100~pc in diameter, would not be
resolved. Although it is clear that nuclear rings with diameters
smaller than a few hundred parsec exist, it is unclear at present how
common such rings are. Because it is also unknown whether such small
rings occur preferentially in later- or earlier-type spirals, we show
in Fig.~\ref{ringtype} how the relative size of the nuclear rings
identified here relates to the morphological type of the host galaxy
(similar results are obtained when plotting the physical or projected
diameters of rings, in kpc or arcsec, but are not shown here).  There
is no evidence for a trend, although the three largest rings do occur
in galaxies of later types. Buta \& Crocker (1993) show a similar
figure in their paper and report a ``hint of a weak type dependence'',
but in the sense that smaller rings occur in galaxies with later
types.

As seen in Figs.~\ref{typecircumnucl} and~\ref{typeRatioCN}, nuclear
rings occur predominantly in host galaxies with intermediate types,
with a preference for type~4, or Sbc galaxies. We find no nuclear
rings in galaxies with types later than 6 (Scd).  This can be
understood within the framework of formation of nuclear rings as a
result of gas accumulation near dynamical resonances set up in the
disk between stellar orbital precession rate and the bar pattern
speed.  For this to occur, a sizeable bulge is needed, which is mostly
absent from galaxies of the latest types, thus explaining the absence
of nuclear rings in these galaxies.  Another prerequisite of nuclear
rings, the availability of gas, may not be fulfilled in many
early-type galaxies, thus explaining the slightly lower nuclear ring
fractions in those galaxies compared to, e.g., those of type Sbc.

\subsection{Nuclear rings in non-barred hosts?}

\begin{figure}
\psfig{figure=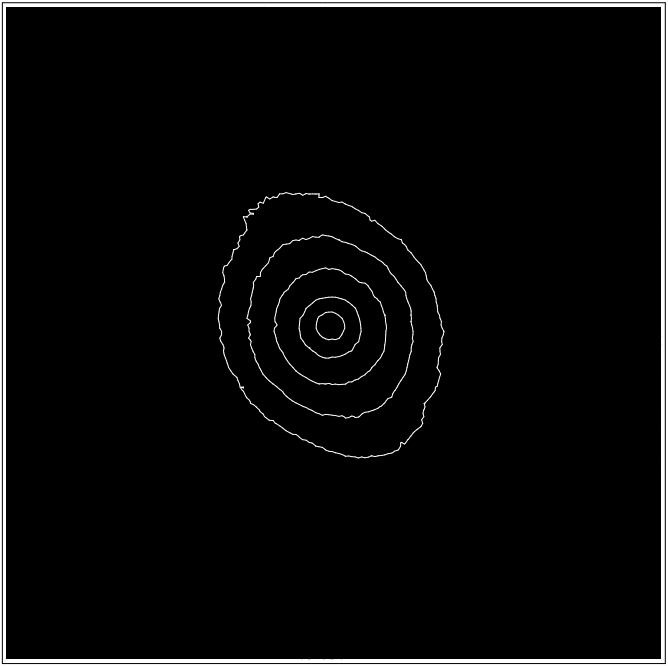,angle=0,width=9cm}
\caption{$K_{\rm s}$ image of the central region of NGC~4736, showing
the small bar, of radius $\sim10$\sec\ (200~pc), most clearly in the
outermost white contour.  North is up and East to the left, and the
scale is indicated by the black bar in the bottom left corner, which
represents a length of 10 arcsec, or about 200~pc in this galaxy.
Contours are shown at surface brightness levels of 16 to 12\,mag
arcsec$^{-2}$, in steps of 0.5\,mag arcsec$^{-2}$.  Based upon data
from Paper~I.}
\label{N4736fig}
\end{figure}

Two nuclear rings occur in galaxies which are classified as non-barred
(SA) in the RC3, namely NGC~1068 and NGC~4736.  It is well known that
NGC~1068 hosts a relatively small bar (of radius 14 arcsec or about
1~kpc; Laine et al.  2002) which is easily visible in the NIR, yet
hardly at all at optical wavelengths (Scoville et al. 1988; Thronson
et al.  1989).  NGC~4736 is a similar case, where a small nuclear bar
is seen in optical and especially NIR images (Shaw et al.  1993;
M\"ollenhoff et al.  1995; Wong \& Blitz 2000; Fig.~\ref{N4736fig}).
In Fig.~\ref{N4736fig}, the maximum elongation of the isophotes is
seen to occur at a radius of about 10 arcsec or 200~pc, although
M\"ollenhoff et al.  model the bar to be slightly larger. In addition,
the disk of NGC~4736 is ovally distorted (Bosma et al.  1977;
M\"ollenhoff et al. 1995; Wong \& Blitz 2000).  We can thus state that
all our nuclear rings occur in barred galaxies, which confirms the
current interpretation where the rings form as a result of gas
accumulation near one or more ILRs driven by a bar.

Although a direct link between nuclear rings and barred host galaxies
can be traced back to the early work of, e.g., S\'ersic \& Pastoriza
(1967), rings of all types (nuclear, inner, outer) do occur in
galaxies classified as unbarred, but as Buta \& Combes (1996) indicate
in their review, some non-axisymmetry in the gravitational potential
can usually be identified, either due to a small or otherwise not
obvious bar (e.g., NGC~1068 and NGC~4736, discussed above), an oval,
or an interaction. As an example we mention the nuclear ring in the
small galaxy NGC~278, which is not only classed as unbarred, but which
also shows no sign at all of a large or nuclear bar in high-quality
optical and NIR images.  There is, however, evidence from \hi\
observations for a recent interaction and minor merger, and this event
is postulated to be the cause of a non-axisymmetry in the
gravitational potential, which in turn causes the nuclear ring (Knapen
et al. 2004b).

\subsection{Large nuclear rings occur in weak bars}

\begin{figure}
\psfig{figure=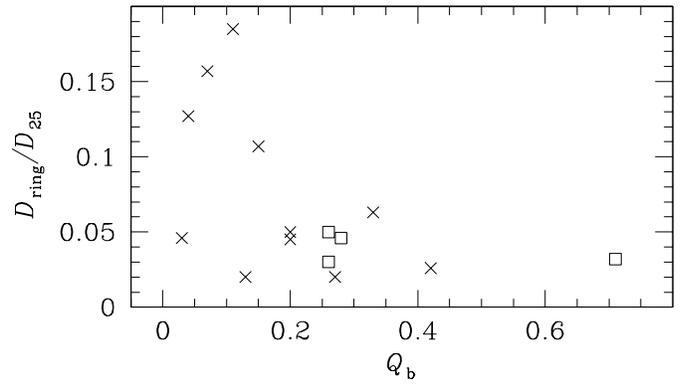,angle=0,width=9cm}
\caption{Relative size,  or ring diameter divided  by galaxy diameter,
of the nuclear rings identified  in our sample (crosses) as a function
of  the gravitational  torque ($Q_{\rm  b}$) of  the bar  of  its host
galaxy.   For one  host  galaxy  (NGC~4736) no  $Q_{\rm  b}$ has  been
published. With a relative nuclear ring  size of 0.152 and a weak bar,
the nuclear ring  in this galaxy would, however, be  in the upper left
corner.  Four galaxies from Knapen et  al. (2002) which are not in our
current sample have also been plotted (squares).}
\label{ringdiam}
\end{figure}

Knapen et al.  (2002) used measurements of the relative sizes of
nuclear rings to confirm observationally the expectation from theory
and modelling (Knapen et al.  1995; Heller \& Shlosman 1996) that
stronger bars host smaller nuclear rings.  This result confirms the
expectation that nuclear rings, being gaseous features, can only occur
in those regions where the main orbit family in the bar (that of the
so-called $x1$ orbits) does not intersect with the second one, of $x2$
orbits.  In the case of stronger, or thinner, bars, the gas will thus
have to settle further in, and any nuclear rings must be smaller (see
fig.~11 of Knapen et al.  1995).  When plotting the relative sizes
(ring diameter divided by the diameter of their host galaxy) of 10
nuclear rings, Knapen et al. (2002) indeed found this (their
fig.~8). With the 12 nuclear rings of the current paper, plus four
from Knapen et al.  (2002) which are not in our current sample, we can
now check the result found before with more nuclear rings, spanning a
wider range in both relative sizes and bar strengths.

The result of this exercise is shown in Fig.~\ref{ringdiam}, where we
plot the relative size of the nuclear rings, defined here as the
diameter of the ring, as measured from our \ha\ images and tabulated
in Table~\ref{resultstab}, divided by the diameter of the complete
disk of its host galaxy ($D_{25}$ from the RC3), against the
gravitational torque $Q_{\rm b}$ due to the bar.  Values of $Q_{\rm
b}$ have been taken mostly from Block et al. (2001), who analysed the
same sample as analysed in this paper, but where other measurements
were available from the literature (Buta \& Block 2001; Laurikainen \&
Salo 2002) averages were used.  For NGC~1530 and NGC~6951 (a nuclear
ring galaxy from Knapen et al. 2002 included in Fig.~\ref{ringdiam})
new values for $Q_{\rm b}$ were derived by Buta et al.  (2003) and
Block et al.  (2004) using a refinement of the original $Q_{\rm b}$
technique, which allows the separate contributions from spiral arms
and bar to the overall gravitational torque to be quantified.  As a
conservative estimate, we can state that typical uncertainties in
$Q_{\rm b}$ are less than 0.1, and in the relative ring size, less
than 10\%.

Figure~\ref{ringdiam}, then, clearly confirms the rather more
preliminary results presented by Knapen et al. (2002), of stronger
bars hosting smaller nuclear rings. In fact, we can now refine this
conclusion to state that whereas weak bars can host nuclear rings of
any size (with an upper limit, by definition, corresponding to the
radius of 2~kpc which we assumed as the limit to the circumnuclear
region), strong bars can only host small nuclear rings. As summarised
already in the first paragraph of this Section, this is in agreement
with the predictions following from earlier numerical and theoretical
interpretation of the formation and evolution of nuclear rings in
relation to resonances and orbit families in bars (Knapen et al. 1995;
Heller \& Shlosman 1996).

\subsection{Inner rings}

In addition to the nuclear rings discussed above, the images of the
sample galaxies also show a number of well-defined inner rings.  A
non-exhaustive but illustrative list of such rings includes the ones
in NGC~2775, of 82~arcsec or 6.7~kpc in radius, in NGC~3368
(150~arcsec or 5.9~kpc), in NGC~4725 (300~arcsec or 18~kpc), or in
NGC~5850 (140~arcsec or 18.9~kpc).  A study of these inner rings is
outside the scope of this paper.  We note in this context, though,
that we classified the well-known ring in NGC~4736 (e.g., Sandage
1961; Lynds 1974; Pogge 1989b) as a nuclear ring, based on its small
physical diameter (of 2.1~kpc, using an image scale derived from the
Tully 1988 distance of 4.3~Mpc to the galaxy), even though due to the
small distance the angular diameter of the ring is rather large (102
arcsec) and it has historically often been referred to as an inner
ring (e.g., van der Kruit 1976; Bosma et al. 1977; M\"ollenhoff et
al. 1995; Martin \& Belley 1997).

\section{Links to non-stellar activity}

\begin{table*}
\centering
\begin{tabular}{lccccc}
\hline
 & Total & AGN & \hii/SB & non-AGN & non-AGN non-SB\\
\hline
\multicolumn{6}{c}{Nuclear emission}\\
Sample size & 51 & 18 & 6 & 33 & 27 \\
Peak & 26 (51\%) & 13 (72\%) & 5 (83\%) & 13 (39\%) & 8 (30\%)\\
Peak + amorph & 8 (16\%) & 2 (11\%) & 1 ( 17\%) & 6 (18\%) & 5 (19\%)\\
Total peak$^1$ & 34 ( 67\%) & 15 (83\%) & 6 (100\%) & 19 (58\%) & 13 (48\%)\\
None & 11 (22\%) & -- & -- & 11 (33\%) & 11 (41\%)\\
Amorph & 5 (10\%) & 2 (11\%) & -- & 3 (9\%) & 3 (11\%)\\
Stellar + amorph & 1 (2\%) & 1 (6\%) & -- & -- & --\\
\multicolumn{6}{c}{Circumnuclear emission}\\
Sample size & 57 & 21 & 6 & 36 & 30\\
Patchy & 32 (56\%) & 8 (38\%) & 4 (67\%) & 24 (67\%) & 20 (67\%)\\
None & 10 (18\%) & 4 (19\%) & -- & 6 (17\%) & 6 (20\%)\\
Ring & 12 (21\%) & 8 (38\%) & 2 (33\%) & 4 (11\%) & 2 (7\%)\\
Diffuse & 3 (5\%) & 1 (5\%) & -- & 2 (6\%) & 2 (7\%)\\
\hline
\end{tabular}
\caption{Nuclear  and   circumnuclear  \ha\  emission   for  different
categories of  nuclear activity, basically  AGN (including LINER-type
activity) and starburst. Column~2: total sample.  Column~3: AGN, which
includes all  galaxies classified in  NED as either Seyfert  or LINER.
Column~4:  \hii/SB, all  galaxies classified  by NED  as  starburst or
\hii.    Column~5:  All   galaxies  not   classified  as   Seyfert  or
Liner. Column~6:  as column~5,  but now also  excluding the  galaxies of
column~4.  Percentages add up vertically  in the Table, i.e., they refer
to the  fraction of galaxies in  a certain category  (e.g., AGN) which
exhibit a  particular nuclear or circumnuclear  \ha\ morphology (e.g.,
patchy).   Note: $^1$  ``Total peak''  category  is the  total of  the
``Peak'' and ``Peak + amorph'' categories.}
\label{agntab}
\end{table*}

\begin{figure}
\psfig{figure=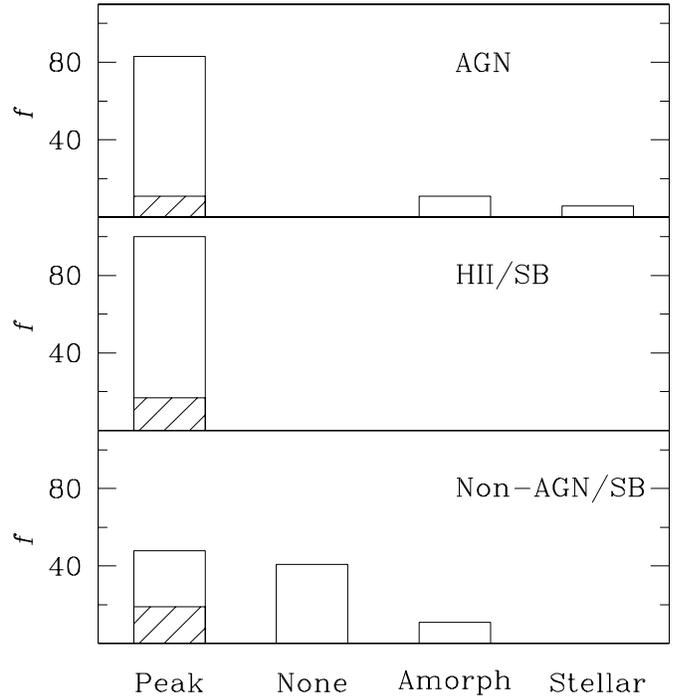,angle=0,width=9cm}
\caption{Distribution of nuclear \ha\ emission morphology in those
galaxies classified as AGN (including both Seyfert and LINER) in NED
({\it top panel}), those classified as starburst or \hii\ ({\it middle
panel}), and those galaxies classified as neither ({\it lower panel}).
Fractions are given as percentages of the total numbers of galaxies in
each category (AGN, etc).  The ``peak'' category denotes the total of
``peak'' and ``peak+amorph'', but the shaded regions identify those
galaxies with a ``peak+amorph'' nuclear \ha\ morphology.}
\label{agnnucl}
\end{figure}

\begin{figure}
\psfig{figure=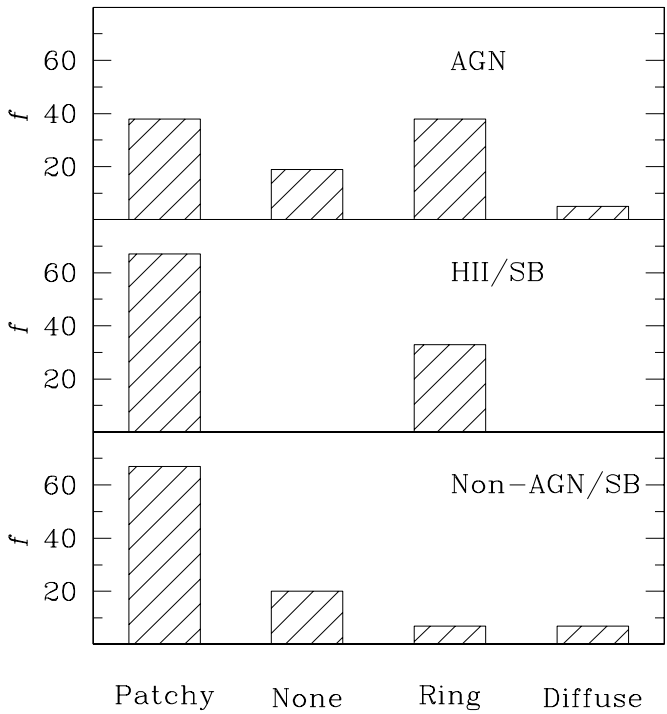,angle=0,width=9cm}
\caption{As Fig.~\ref{agnnucl}, now for circumnuclear \ha\ emission 
morphology.}
\label{agncnucl}
\end{figure}

The nuclear and circumnuclear \ha\ morphologies of the sample galaxies
as classed  by their nuclear activity is  shown in Table~\ref{agntab},
and shown  in Figs.~\ref{agnnucl} for the  nuclear, and \ref{agncnucl}
for the circumnuclear morphology  (where fractions rather than numbers
are now plotted in the histograms). The different classes were defined
on the basis of the galaxy classifications in the NED, as follows.  We
class as ``AGN'' all those  galaxies where the NED classes galaxies as
either Seyfert (of any  type) or LINER (including classifications such
as LINER/starburst). We thus place  all galaxies in this bin for which
there is  evidence for non-stellar nuclear  activity.  The ``\hii/SB''
category contains all those galaxies  either classified as \hii\ or as
starburst in  the NED. The final classes,  non-AGN and non-AGN/non-SB,
were derived  from the  previous two classes,  where the  latter class
contains those galaxies  for which the NED does  not list any evidence
for either AGN or starburst  activity. With these classes, we can thus
make a rough distinction between AGN, starbursts, and galaxies without
any nuclear  activity.  The NED classification of  the sample galaxies
has been tabulated in Paper~I.

To check the NED classification, we correlated our sample with that of
Ho et al.  (1997), who give nuclear activity classes for over 450
galaxies based on their own detailed spectroscopy.  We find that for
the 43 galaxies in both samples the classifications (for our purpose,
thus in broad categories ``AGN'' and ``starburst'') by Ho et al.  and
as obtained from the NED agree well, with disagreements only in less
than 10\% of the cases.  We thus feel confident to use the information
from the NED, but are aware of the caveat that where no spectroscopic
information of a galaxy's nucleus is available from the literature,
the galaxy would automatically be classed as ``non-AGN,
non-starburst'' in our analysis.  Given, however, that our sample
galaxies are nearby, bright, and generally well-studied, we assume
that only a tiny minority of galaxies, if any at all, can fall in this
category.

The results for {\it nuclear} morphology (Fig.~\ref{agnnucl}) are not
surprising: most AGN are centrally peaked in \ha, {\it all} starbursts
are, and most of the non-peaked morphologies are found among the
non-AGN/non-SB galaxies.  Another result worth reiterating in this
context is that even among the non-AGN/non-SB galaxies, almost half
have peaked nuclear \ha\ emission.

The AGN and the rings lead to a very interesting {\it circumnuclear}
\ha\ emission result (Fig.~\ref{agncnucl}).  Whereas more than half
(56\%) of all galaxies have patchy circumnuclear emission, only 38\%
of the AGN have. An equally high fraction though (38\% of the AGN
hosts) have nuclear rings.  Less surprisingly, the starbursts all have
patchy (four out of six) or ring (the remaining two) \ha\ morphology.

Of the 12 nuclear rings we identified, only two (17\% of nuclear
rings) occur in a galaxy which hosts neither a starburst nor an AGN,
and the great majority (8/12 or 67\% of the rings) occurs in AGN
hosts.  This result, and the one described in the previous paragraph,
strongly indicate that there is a direct statistical link between
nuclear rings and non-stellar nuclear activity. This link has, to our
knowledge, not been indicated before, and is interesting because the
spatial scales involved in either of these processes are rather
different: a few 100~parsec to a kiloparsec for the nuclear rings,
versus much smaller scales, possibly of the order of AUs, for the
black hole accretion powering the AGN.  Yet the link between
star-forming nuclear rings, regarded as prime indicators of very
recent gas inflow, and AGN indicates that recent gas inflow on scales
of up to a kpc may have more to do with the AGN as other studies, for
instance of bar fractions in AGN and non-AGN (Knapen et al. 2000;
Laine et al. 2002), would lead one to believe.  We note the effect and
its possible implications here, but point out the need to confirm it
through the study of larger samples of nuclear rings. We also point
out a possibly related result reported in the literature, namely that
of correlations between inner and outer rings in galaxies and the
occurrence of AGN activity, e.g. by Arsenault (1989) and by Hunt \&
Malkan (1999), which are intriguing if only because the spatial scales
of those rings are even further removed from typical AGN scales.

\section{Conclusions}

From a complete set of \ha\ images of a carefully selected sample of
57 relatively large, Northern spiral galaxies with low inclination, we
study the distribution of the \ha\ emission in the nuclear and
circumnuclear regions, emission which traces predominantly massive
young stars.  The images form part of a larger set comprising also
NIR $K_{\rm s}$ and optical broad-band ($B$, $R$ and $I$) images
of all galaxies (Papers I, II). The majority of the \ha\ images have
been taken with the 1~m JKT on La Palma over the past few years, at a
spatial resolution of around 1.5~arcsec, or around 100 parsec. The
main results described in the current paper can be summarised as
follows.

\begin{itemize}

\item The {\it nuclear}  \ha\  emission is peaked in  most  of the  
sample galaxies, and significantly more often so among AGN host galaxies
than among non-AGN.

\item The {\it circumnuclear} \ha\ emission, defined to be that
originating within  a radius of two kpc,  is often patchy  in late-type,
and absent or in the form of a nuclear ring in early-type galaxies.

\item There  is no  clear correlation  of  nuclear or
circumnuclear  \ha\ morphology with the presence  or absence of a bar in
the host  galaxy, {\it except} for the  nuclear rings which all occur in
barred hosts.

\item The presence or absence of close bright {\it companion galaxies}
does not affect the circumnuclear \ha\ morphology, but their presence
does correlate with a higher fraction of nuclear \ha\ peaks.

\item Star-forming {\it nuclear rings} occur in one of every five
galaxies (21\%$\pm$5\%), although at the rather low spatial resolution
in our images this number must be considered a lower limit.

\item We  confirm earlier results that only weaker bars can host  larger 
nuclear rings, an effect which is  explained in terms of the predominant
orbit families in bars and their relation to rings.

\item The nuclear rings identified here occur predominantly  in galaxies 
also hosting  an AGN, and only two  of our 12  nuclear rings occur in a
galaxy which  is neither an  AGN nor a   starburst host.  This indicates
some link between nuclear  rings  and non-stellar nuclear activity,  and
since nuclear  rings are prime  indicators of  very  recent gas  inflow,
common fuelling processes may be involved.

\end{itemize}

\begin{acknowledgements}

I thank my collaborators on Paper~II, Tom Bradley, Daniel Bramich,
Stuart Folkes and Sharon Stedman, for their help in producing the \ha\
images used in the current paper.  Torsten B\"oker and Isaac Shlosman
are acknowledged for helpful comments. We used the HyperLEDA database,
hosted by the Observatoire de Lyon (http://leda.univ-lyon1.fr/). This
research has made use of the NASA/IPAC Extragalactic Database (NED)
which is operated by the Jet Propulsion Laboratory, California
Institute of Technology, under contract with the National Aeronautics
and Space Administration.

\end{acknowledgements}

\label{lastpage}

\end{document}